\RequirePackage{fix-cm}
\documentclass[twocolumn,epjc3]{svjour3}  
\smartqed  % flush right qed marks, e.g. at end of proof
\RequirePackage{graphicx}
\usepackage{color}
\usepackage{amssymb,bbold}
\usepackage{graphicx,color}
\usepackage[usenames,dvipsnames]{xcolor}
\usepackage[section]{placeins}
\usepackage{amsmath}
\usepackage{subfigure}
\usepackage[normalem]{ulem}
\usepackage{booktabs}
\usepackage{xcolor}
\usepackage{epsfig,graphics,color,graphicx,amsmath}
\colorlet{darkgreen}{green!50!black}
\colorlet{brightyellow}{yellow!75!red}
\colorlet{orange}{red!50!yellow}
\colorlet{darkblue}{blue!60!black}
\colorlet{darkred}{red!80!black}
\usepackage{widetext}
\def\bwt{\begin{widetext}}
\def\ewt{\end{widetext}}
\usepackage{soul}

\usepackage{mathtools}
\usepackage{mathrsfs}
\usepackage{bm}
\usepackage{relsize}
\usepackage{indentfirst}

\usepackage{xcolor}

\usepackage{amssymb,amsmath,multirow,epsfig,graphicx,color}

\usepackage{epsfig}
\usepackage{graphicx,amsmath}
\def\be{\begin{eqnarray} &&}

\def\ee{\end{eqnarray}}

\begin{document}

\title{Quark Stars in $D_3$-$D_7$ Holographic Model }
\author{M.~Aleixo \thanksref{e1,addr1} \and C.H. Lenzi \thanksref{e2,addr1} \and W. de Paula \thanksref{e3,addr1} \and R. da Rocha \thanksref{e4,addr2}}
\institute{Instituto Tecnol\'ogico de Aeron\'autica,  DCTA,
12228-900 S\~ao Jos\'e dos Campos,~Brazil \label{addr1} \and  Federal University of ABC, Center of Mathematics, Santo Andr\'e, 09580-210, Brazil \label{addr2}}
\thankstext{e1}{e-mail: s.michaelaleixo@gmail.com} 
\thankstext{e2}{e-mail: chlenzi@ita.br}
\thankstext{e3}{e-mail: wayne@ita.br} 
\thankstext{e4}{e-mail: roldao.rocha@ufabc.edu.br}

\date{\today}
\maketitle
\begin{abstract}
This work investigates static and dynamical quark star properties within a $D_3-D_7$ holographic model. We solve the Tolman-Oppenheimer-Volkoff equations for the quark matter equation of state obtained from the brane configuration. We determine the mass-radius diagram
for a range of model parameters and compare with recent NICER observational data for the pulsars PSR J$0030+0451$ and PSR J$0740+6620$. Motivated by the GW170817 event detected by the LIGO-Virgo collaboration, we also calculate the tidal deformability parameter obtained for each component of the binary star system. We show that quark stars composed of flavor-independent quark matter derived from the $D_3-D_7$ holographic model are not able to satisfy simultaneously the LIGO-Virgo and NICER astrophysical bounds.

\end{abstract}

%\begin{keyword} 
%Holographic models, Compact Stars, Quark Stars.
%\end{keyword} 

\section{ Introduction}

The detection of the gravitational waves (GW) \cite{LIGOScientific:2017ync} and Gamma-ray burst (GRB) \cite{LIGOScientific:2017zic} from a binary neutron star (NS) merger, the GW170817 event, brought new valuable information for the description of compact star properties. In particular, the details of the NS structure become more relevant as the separation between each binary companion decreases \cite{LIGOScientific:2017vwq}. In this context, the tidal deformability extracted from the GW170817 data \cite{LIGOScientific:2017vwq,LIGOScientific:2018cki,LIGOScientific:2018mvr} gives new dynamical constraints for NS models. 

Understanding the composition of the NS interior is an important astrophysical open problem \cite{Lattimer:2015nhk}. In their inner core, which is believed to achieve very high densities, few times the nuclear saturation density, theoretical models predict the existence of hyperons \cite{Glendenning:1991es,Bombaci:2008wg,Dexheimer:2008ax} or deconfined quark matter \cite{Bodmer:1971we,Witten:1984rs,Terazawa:1978ni,Lattimer:2006xb}. Indeed, there are also indirect observational shreds of evidence that open the possibility of forming stable compact stars only with quark matter, known as quark stars (QS), which can play the role of laboratories to investigate the very fundamental physics underlying systems at supranuclear densities, under strong gravitational fields \cite{Bombaci:1997zz,Cheng:1998qc,Li:1999wt,Li:1999mk,Burgio:2011wt}. Therefore, exploring the possibility of a description of a NS with exotic content, or being a core of quark matter in hybrid stars \cite{Alford:2004pf,Pereira:2017rmp,Blaschke:2022egm,Lobato:2020fxt,Lenzi:2022ypb} or QS \cite{Haensel:1986qb,Xu:2003xe,Lugones:2015bya,Lourenco:2021lpn,Chu:2023rty} is an active area of study.

The AdS/CFT correspondence allows to treat strongly-coupled quantum systems in terms of gravitational duals \cite{Maldacena:1997re}. There are applications of such proposal in many areas, from condensed matter systems \cite{Sachdev:2010ch} to the description of the quark-gluon plasma (QGP) produced in experiments of heavy-ions collisions \cite{Policastro:2001yc,Brambilla:2014jmp}. In particular, it is worth mentioning how close to experimental data \cite{Bernhard:2019bmu} is the prediction of the shear viscosity-to-entropy ratio of the QGP from holographic models, which attains the lowest value among any kind of matter in Nature, the nearest to the Kovtun-Son-Starinets limit \cite{Kovtun:2004de}. The original duality maps the generating functional of the correlation functions of $\mathcal{N} = 4$ super Yang-Mills (SYM) theory in 4D flat space to partition functions of type IIB string theory in AdS$_{5} \times$ S$^{5}$\cite{Witten:1998qj}. Within the holographic concept, there are many attempts to incorporate some features of quantum chromodynamics (QCD), such as confinement, chiral symmetry breaking, and the hadronic spectrum, besides the phase structure at large baryon-chemical potentials, and the equation of state governing high-density regimes, as the ones expected to take place in the quarkyonic matter core of NS \cite{Klebanov:2000hb,Klebanov:2000nc,Maldacena:2000yy,Karch:2006pv,dePaula:2008fp,Bianchi:2010cy,dePaula:2009za,Ballon-Bayona:2023zal}. 

Here we are mainly interested in the description of dense QCD matter for the analysis of the QS properties. For this end, we focus on the $D_3-D_7$ system \cite{Karch:2002sh}, where a configuration of $N_c$ $D_3$ branes and $N_{f}$ $D_7$ probe branes are considered\footnote{$N_c$ and $N_f$ are the number of colors and flavors, respectively.}. By taking the 't Hooft limit, $N_c\to \infty$, $g_s \to 0$ with $\lambda = g^2_s \, N_c$ fixed and large, in the near-horizon limit of $D_3$ branes, one obtains $AdS_5 \times S^5$ with the $N_f$ $D_7$-branes wrapping $AdS_5 \times S^3$ \cite{Karch:2007br}. The presence of the $D_7$ probe brane generates new degrees of freedom, whose low-energy dynamics are described by the Dirac-Born-Infeld (DBI) one, in $AdS_5 \times S^3$, where the time component of the U(1) gauge field is dual to the chemical potential $\mu$. These degrees of freedom correspond to open string fluctuations on the $D_7$-brane. The asymptotic distance between the $D_3$ and $D_7$-branes is a mass parameter $m$, which, in this context, is interpreted as the constituent quark mass \cite{Hoyos:2021uff}. This ulterior open-open string duality maps operators of mesonic type, in the conformal field theory, to $D_7$-brane fluctuations, on the gravitational sector, additionally to the original AdS/CFT, whose gravity is regulated by the near-horizon geometry of $D_3$-branes.  Gauge-invariant field theory bilinear operators are, in this way, dual objects mapped to fluctuations of the $D_7$ probe brane living in the  AdS$_5 \times S^5$ compactified space.

Considering the grand canonical ensemble, one can study the thermodynamic properties of the model, as implemented in Refs. \cite{Karch:2007br,Hoyos:2021uff,Mateos:2006nu,Kobayashi:2006sb,Mateos:2007vn,Karch:2009zz,Nakamura:2007nx,Erdmenger:2008yj,Ammon:2008fc,Basu:2008bh}. The proposal regards obtaining the equation of state (EOS) for zero temperature of such holographic model and, with the use of the Tolman-Oppenheimer-Volkoff (TOV) equation for the hydrostatic equilibrium, to analyze static and dynamical properties of QS. There is a vast literature where holographic concepts were used to discuss compact stars, as reported by Refs. \cite{Hoyos:2021uff,Hoyos:2016zke,Annala:2017tqz,BitaghsirFadafan:2019ofb,BitaghsirFadafan:2020otb,Mamani:2020pks,daRocha:2017cxu,Meert:2020sqv,daRocha:2021aww,Most:2021zvc,Kovensky:2021kzl,Demircik:2021zll} and  references therein.

In what follows, we will obtain the free energy of the flavor fields,  decoupled from the adjoint fields. After determining the holographic EOS for the quark matter, we calculate the mass distribution profile and the mass-radius diagram in terms of the constituent quark mass $m$. By varying the parameter $m$, we compare the results with the observational data analysis of the Neutron Star Interior Composition Explorer (NICER) on the values of mass and radius of the massive pulsars PSR J$0030+0451$ \cite{Riley:2019yda,Miller:2019cac} and PSR J$0740+6620$ \cite{Miller:2021qha,Riley:2021pdl}. Finally, we consider an NS merger and compare the tidal deformability obtained in the holographic model with the data that comes from the LIGO-VIRGO Collaboration on the event GW170817 \cite{LIGOScientific:2017ync}. 
It is worth mentioning that similar analyses were performed in \cite{Annala:2017tqz}, where it was presented different compact stars solutions. For each solution, relevant quantities derived from the holographic model were compared with Ligo-Virgo observational data. In the present work, we focus on QS properties also taking into account recent NICER observational data for constraining the model.

\section{The holographic model}

In the adopted framework, one considers the 't Hooft limit for the $D_3-D_7$ system, obtaining an $AdS_{5} \times S^{5}$ with the $D_7$-branes wrapping the $AdS_{5} \times S^{3}$ space \cite{Karch:2007br}. The metric reads 
\begin{equation}
    ds^2 = {u^2 \over \mathcal{R}^2} \eta_{\mu\nu} dx^{\mu} dx^{\nu} + {\mathcal{R}^2 \over u^2} \left(d\bar{\rho}^2 + \bar{\rho}^2 d\Omega_{3}^2 + dy^2 + dz^2\right) \, ,
\end{equation}
where $\eta_{\mu\nu}$ is the Minkowski metric in $4$ dimensions and $\mathcal{R}$ is the AdS radius. The holographic coordinate $u$ is written as $u^2 = \bar{\rho}^2 + y^2 + z^2$ and the coordinates $\bar{\rho}$ and $\Omega_3$ belong to the $D_7$ brane world volume. 
The DBI action has the form
\begin{align}
    S_{D_{7}} = - N_{f} \, T_{D_{7}} \, \int d^{8} \xi \, e^{- \phi} \, \sqrt{- \, \det(g + 2 \pi \, \alpha' \, F)} \, ,
\end{align}
where $T_{D_{7}}$ is the tension of the $D_7$-brane, $g$ is the induced metric on the $D_7$ worldvolume, the AdS radius was set to one, $\phi$ is the dilaton field, $\alpha'$ is the inverse of the string tension and $F$ is the field strength of a $U(1)$ gauge field $A^{\mu}$, whose only non-vanishing component is the temporal one $A_{t}(\bar{\rho})$. 

The DBI Lagrangian can be written as 
\begin{equation}
\mathcal{L}_{DBI} = - \mathcal{N} \, \bar{\rho}^{3} \,  \, \sqrt{1+z'^{2} \,  -  A_{t}'^{2}} \, ,
\end{equation}
where $\mathcal{N} = {\pi^2 \over 2} \, N_{f} \, T_{D_{7}}$. The variation of the Lagrangian with respect to $z$ and $A_{t}$ is zero. Therefore, one has two conserved quantities, $c$ and $d$, respectively given by
\begin{eqnarray}
   c&=& -\frac{1}{\mathcal{N}} \, \frac{\partial \mathcal{L}_{DBI}}{\partial z'} =  \frac{\bar{\rho}^{3} \, z'}{\sqrt{1 + z'^{2} - A_{t}'^{2}}}\, , \\
    d&=&\frac{1}{\mathcal{N}} \, \frac{\partial \mathcal{L}_{DBI}}{\partial A_{t}'} =  \frac{\bar{\rho}^{3} \, A_{t}'}{\sqrt{1 + z'^{2} - A_{t}'^{2}}}\,.
\end{eqnarray}
The holographic dictionary relates the constituent quark mass and the chemical potential $\mu_q$ with the asymptotic boundary of the fields $A_t$ and $z$, specifically, one has $A_{t} (\bar{\rho} \rightarrow \infty) = \, \, \mu_q$ and $z(\bar{\rho} \rightarrow \infty) = \, \, m$. After this identification, one can show that the conserved quantities $c$ and $d$ are related to the physical quantities $\mu_q$ and $m$ \cite{Karch:2007br}. At zero temperature, the thermodynamic potential in the grand canonical ensemble can be obtained from the regulated on-shell action \cite{Karch:2009zz}. The free energy density can be written as \cite{Hoyos:2016zke}
\begin{equation}
\mathcal{F} = \mathcal{F}_{\mathcal{N}=4} + \mathcal{F}_{flavor} \, ,
\label{FreeEnergy}
\end{equation}
where the first part of the r.h.s. in Eq. (\ref{FreeEnergy}) is associated with the color charge and vanishes in the zero temperature limit \cite{Mateos:2006nu}. In this case, the flavor contribution reads \cite{Hoyos:2016zke}
 \begin{equation}
\mathcal{F}_{flavor} = - {3 \over {4 \, \pi^2}} (\mu_{q}^{2} - m^{2})^{2} \, ,
\label{Fenergy}
\end{equation}
where the number of colors and flavors are three and the 't Hooft coupling constant $\lambda$ was chosen to reproduce the Stefan-Boltzmann expression for large density. 
  
\section{ Holographic compact stars}
\label{hcs}
Considering the thermodynamic relation between the pressure and the free energy, $p = - \mathcal{F}_{flavor}$, together with the expression $\varepsilon = \mu_{q}  \, \frac{\partial p}{\partial \mu_{q}} - p$, where $\varepsilon$ is the energy density and the label $q$ is associated to the quark, one obtains the EOS of the holographic model as \cite{Annala:2017tqz}
\begin{equation}
\varepsilon = 3 p + \frac{2 \sqrt{3} \, m^{2} }{\pi} \sqrt{p} \, \, ,
\label{energydensity}
\end{equation}
where $p$ is the pressure. To verify that causality is respected in the model, it is useful to write the explicit expression of the sound velocity $v_{s}$, which is given by 
\begin{equation}
v_{s} = \sqrt{\frac{\partial p}{\partial \varepsilon} }= \sqrt{\frac{\pi \sqrt{p} }{\sqrt{3} \, m^{2} + 3 \pi \sqrt{p}} } \, .
\label{dpde}
\end{equation}

To ensure the hydrostatic equilibrium for a spherically symmetric distribution of mass, one has to solve the TOV equations, written in natural units ($G=c=1$), given by 
\begin{eqnarray}
\frac{dp (r)}{dr} &=& -\frac{ M (r) \, \rho(r)}{r^{2}} \left(1 + \frac{4 \pi r^{3} p(r)}{M (r)} \right) \left(1 + \frac{p(r)}{\varepsilon(r)} \right) \nonumber\\
&& \times \, \left(1 - \frac{2 M (r)}{r} \right)^{-1}, \label{TOV1}\\
   \frac{d M (r)}{dr} &=& 4 \pi r^{2} \rho(r), \label{TOV2}
\end{eqnarray}
where the $M(r)$ is the Misner-Sharp mass inside the radius $r$ and $\rho(r)$ is the mass density.

\section{Tidal deformability}
\label{hcs1}
The LIGO-Virgo collaboration detected GW 
\cite{LIGOScientific:2017ync} and GRB from a binary NS merger \cite{LIGOScientific:2017zic}, the GW170817 event. This system provides valuable information concerning the deformations due to the gravitational interaction between the two involved neutron stars \cite{Damour:2009vw}, which can be given, to linear order, in terms of the dimensionless tidal deformability parameter $\Lambda$ \cite{Hinderer:2007mb}, reading 
\begin{equation}
    \Lambda = \frac{Q_{ij}}{\varepsilon_{ij}} \, ,
\end{equation}
where $Q_{ij}$ is the quadrupole momentum and $\varepsilon_{ij}$ is the tidal field. The induced quadrupole moment is associated with the deformation of a spherically symmetrical object with respect to the flattening of the poles. In terms of the second Love number $k_2$, we have
\begin{equation}
    \Lambda = {2\over 3} \, k_2 \, C^{-5} ,
\label{lambda}
\end{equation}
where $C = M/R$ is the compactness. On a quasi-static regime, the second Love number is given by \cite{Hinderer:2007mb}
\begin{eqnarray}
   k_2 \!& = &\! {8C^5 \over 5} (1 \!-\! 2 C)^2 \left(2 \!+\! 2C(y_R \!-\! 1) \!-\!y_R\right) \nonumber\\
  \!\!\!\!\!\!\!\!\!&& \times \Big\{ 2C \left( 6 - 3 y_R + 3 C (5 y_R - 8)\right) \nonumber\\
  \!\!\!\!\!\!\!\!\!&& + 4 C^3 \left(13 - 11 y_R + C (3 y_R - 2) + 2C^2 (1+y_R)\right) \nonumber\\
 \!\!\!\!\!\!\!\!\! && + 3 (1\!-\!2C)^2 \left(2 \!-\! y_R \!+\! 2 C (y_R \!-\! 1) \right)  \ln{(1\!-\!2C)}\Big\}^{-1} \, , \nonumber\\
\label{k2}
\end{eqnarray}
where $y_R = y(R)$. The function $y(r)$ is a solution of the differential equation $r\,  (dy/dr) + y^2 + y \, F(r) + r^2 \, Q(r) = 0$, with
\begin{eqnarray}
    F(r) &=& {1 - 4 \pi r^2 \left(\varepsilon(r) - p(r) \right) \over g(r)} \, ,\nonumber\\
    G(r) &=& {4 \pi \over g(r)} \left(5 \varepsilon(r) + 9 p(r) + {\varepsilon(r) + p(r) \over v_s^2(r)} - {6 \over 4 \pi r^2} \right) \nonumber\\
    && -4 \left( {m(r) + 4 \pi r^3 p(r) \over r^2  g(r)}\right)^2 \, , \nonumber\\
    g(r) &=& 1 - \frac{2 M (r)}{r} \, .
\end{eqnarray}
In addition, we define the chirp mass parameter $\mathcal{M}$ as
\begin{equation}
    \mathcal{M} \equiv \left(\frac{m_{1}^{3} \, \, m_{2}^{3}}{m_{1} + m_{2}} \right)^{\frac{1}{5}} \, ,
\end{equation}
which is a function of the masses of the two NS companions, $m_1$ and $m_2$. This parameter is relevant to describe the rate of energy transferred away through the gravitational waves. Indeed, the tidal deformability analysis from the observational data of the GW170817 data from LIGO-Virgo is made for a specific value of the system chirp mass \cite{LIGOScientific:2017vwq}.

\section{Results} 
\label{res1}
\begin{figure}[t]
\includegraphics[width=8.5cm, angle=0]{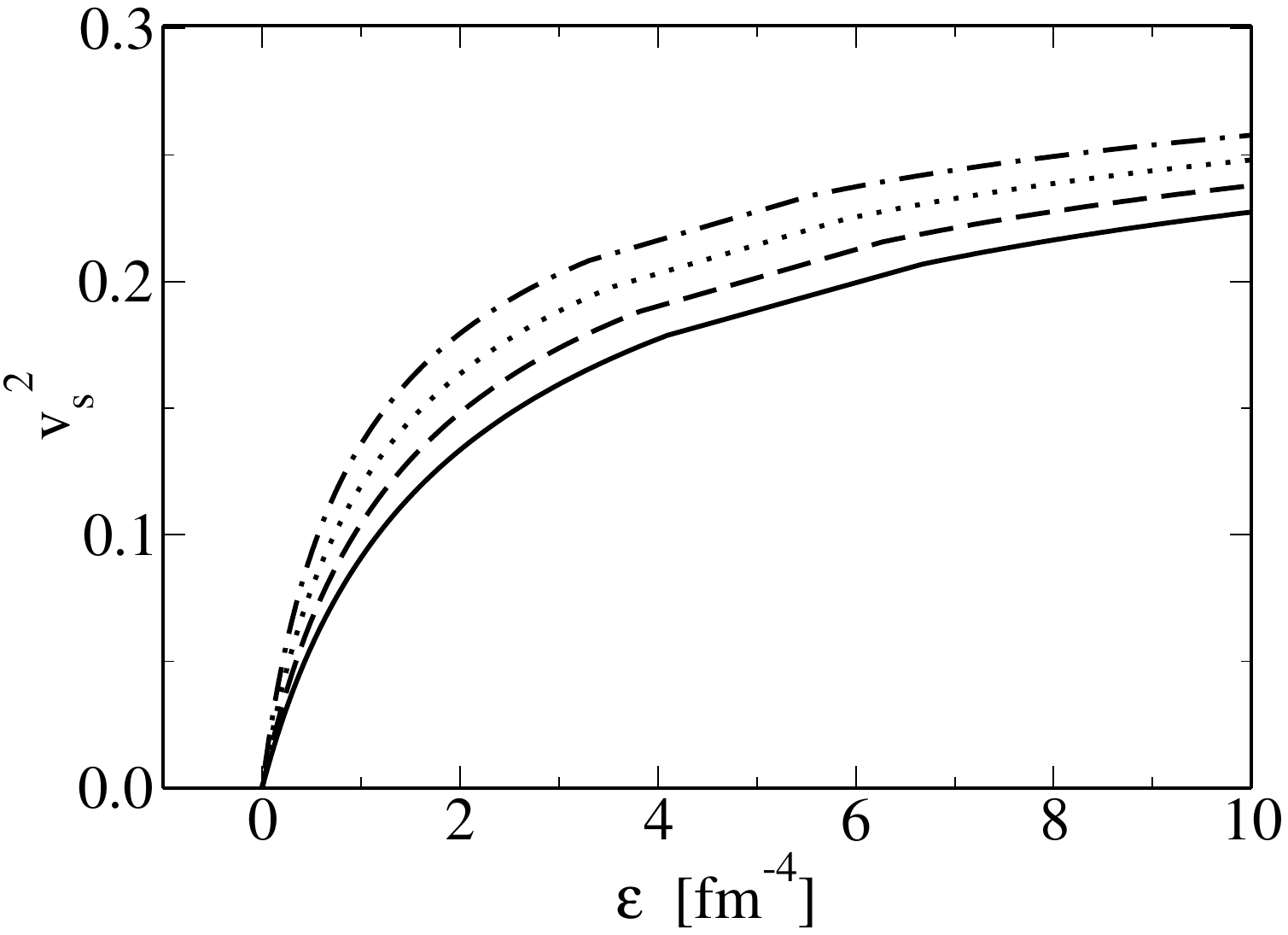}
%\vspace{-0.5cm}
\caption{Sound velocity for each parametrization. Dot-dashed line: $m = 300$ MeV. Dotted line: $m = 320$ MeV. Dashed line: $m = 340$ MeV. Solid line: $m = 360$ MeV.} 
\label{vel_som}
\end{figure}

An important parameter to be analyzed is the speed of sound corresponding to the model. With this information, it is possible to check whether the model does not violate the causality principle ($\partial p /\partial \varepsilon < 1$). Fig.~\ref{vel_som} presents the speed of sound curves, $v_s^2$, as a function of energy density $\varepsilon$. As can be seen, all models do not violate the causality principle.

The solutions of the differential equations system given by Eqs. (\ref{energydensity}), (\ref{TOV1}) and (\ref{TOV2}) has been obtained for constituent quark masses ranging from $m$ = $300$ MeV to $m$ = $360$ MeV. The initial conditions used are $p(0)=p_c$ and $M(0) = 0$, where $p_c$ is the central pressure. The radius $R$ of the star is defined by $p(R)=0$. The outcome is the $M(R)$ sequences of compact stars compatible with the adopted model. The rationale behind the choice of the range of values for $m$ is the following: since $m$ is interpreted as the constituent quark mass, a typical value can be obtained from the infrared value of the quark mass function \cite{Castro:2023bij,Duarte:2022yur}, which value of $345$ MeV was obtained with lattice QCD calculations for the quark propagator \cite{Oliveira:2018lln}. This work proposes to explore a range of values around this number to see if the model can describe observational data of statical and dynamical properties of NS. It will be shown that there is not a parametrization able to satisfy both astrophysical constraints simultaneously.

\begin{figure}[t]
\includegraphics[width=8.5cm, angle=0]{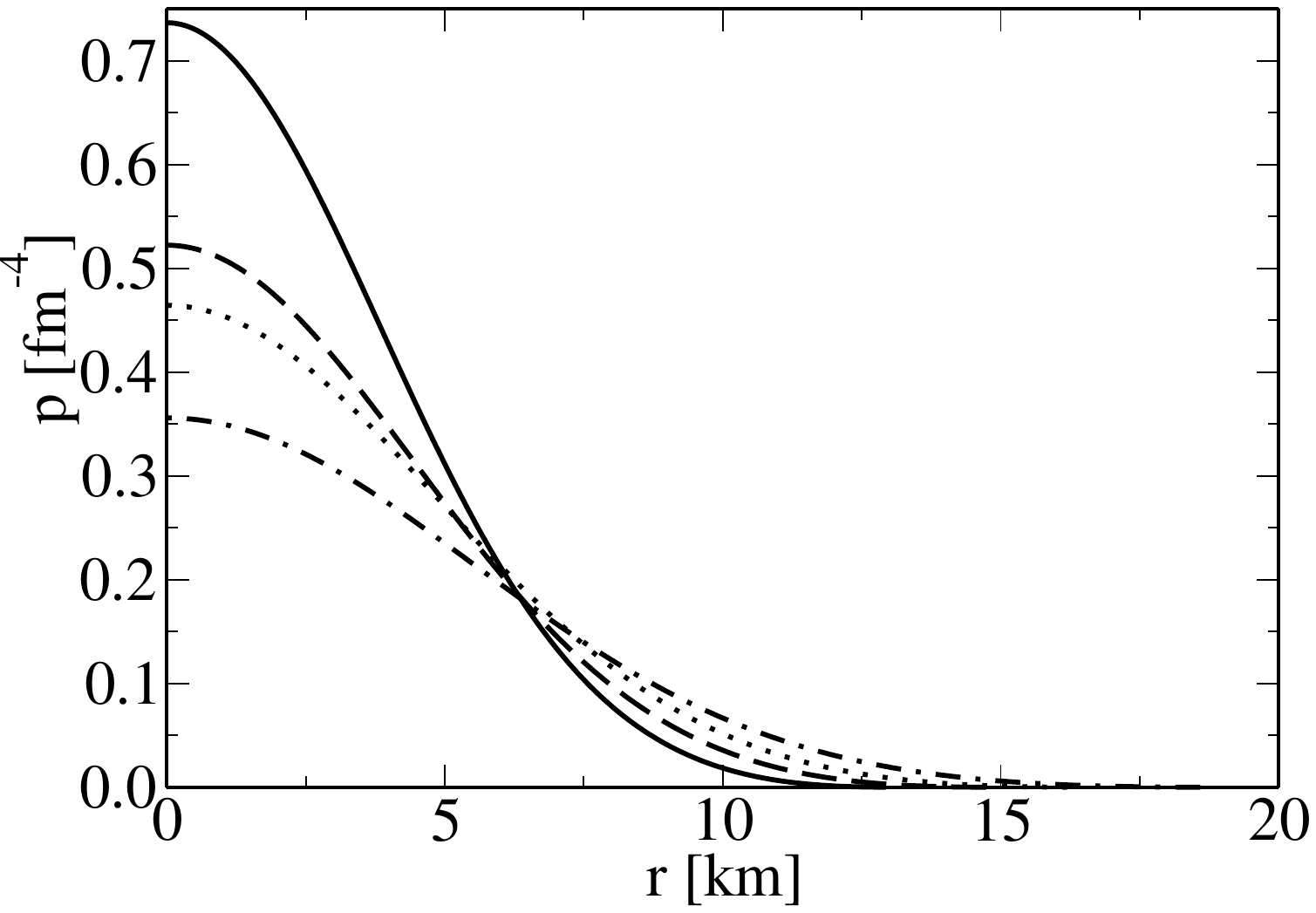}
%\vspace{-0.5cm}
\caption{QS radial profiles for the maximum stellar mass of each parametrization. Pressure versus radial coordinate. Dot-dashed line: $m = 300$ MeV. Dotted line: $m = 320$ MeV. Dashed line: $m = 340$ MeV. Solid line: $m = 360$ MeV.} 
\label{profiles}
\end{figure}

\begin{figure}
\includegraphics[width=8.5cm, angle=0]{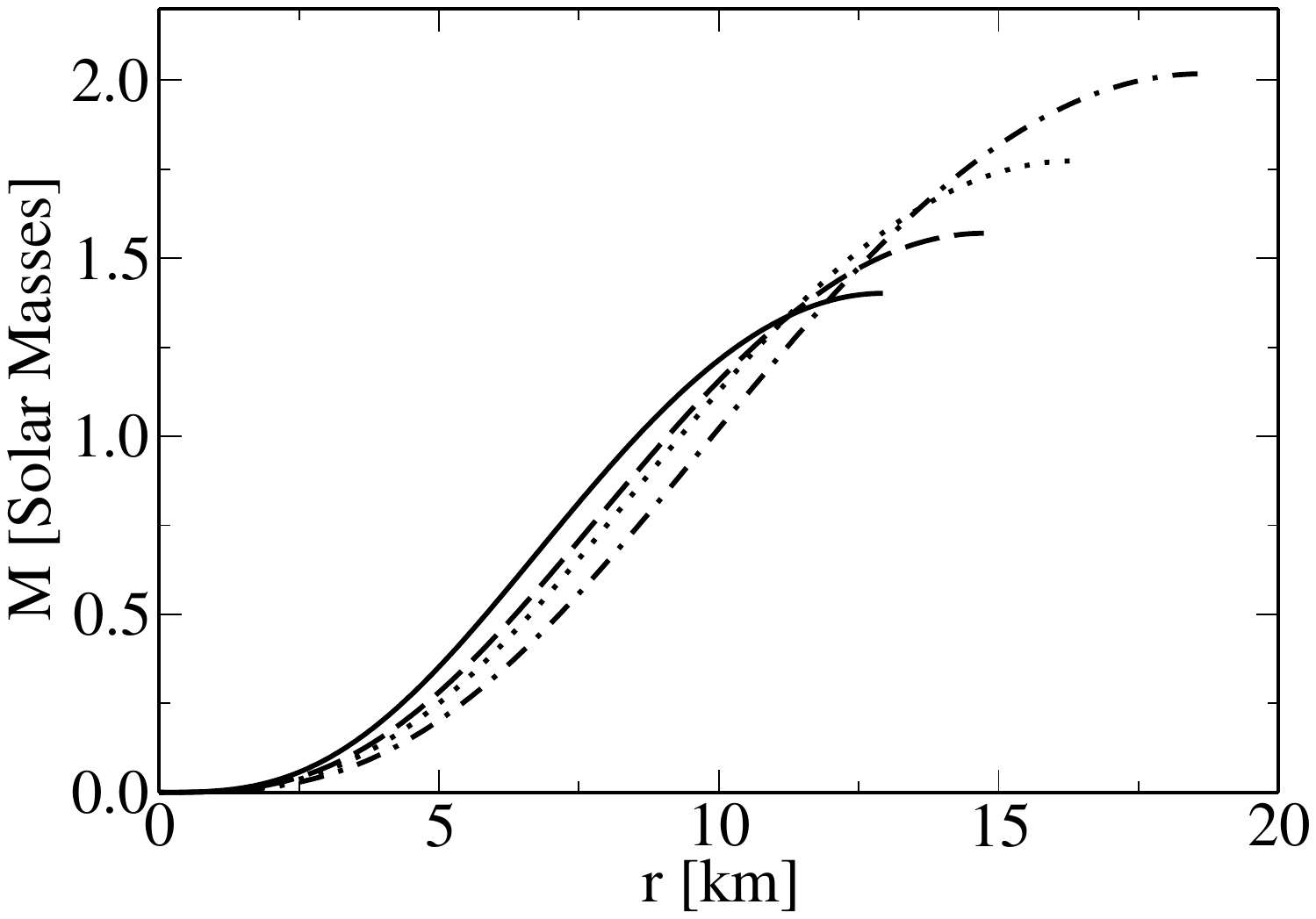}
%\vspace{-0.5cm}
\caption{QS radial profiles for the maximum stellar mass of each parametrization. Mass inside a volume of radius $r$ versus radial coordinate. Dot-dashed line: $m = 300$ MeV. Dotted line: $m = 320$ MeV. Dashed line: $m = 340$ MeV. Solid line: $m = 360$ MeV.} 
\label{profiles1}
\end{figure}

Figs. \ref{profiles} and \ref{profiles1}  present the radial profiles for the maximum star mass of each parametrization. Fig. \ref{profiles} shows that the maximum central pressure is obtained for $m = 360$ MeV, while the minimum is attained for $m = 300$ MeV. Fig. \ref{profiles1} illustrates that the radius of the maximum star mass decreases monotonically with the constituent quark mass.

\begin{figure}
\begin{center}
\includegraphics[width=8.5cm, angle=0]{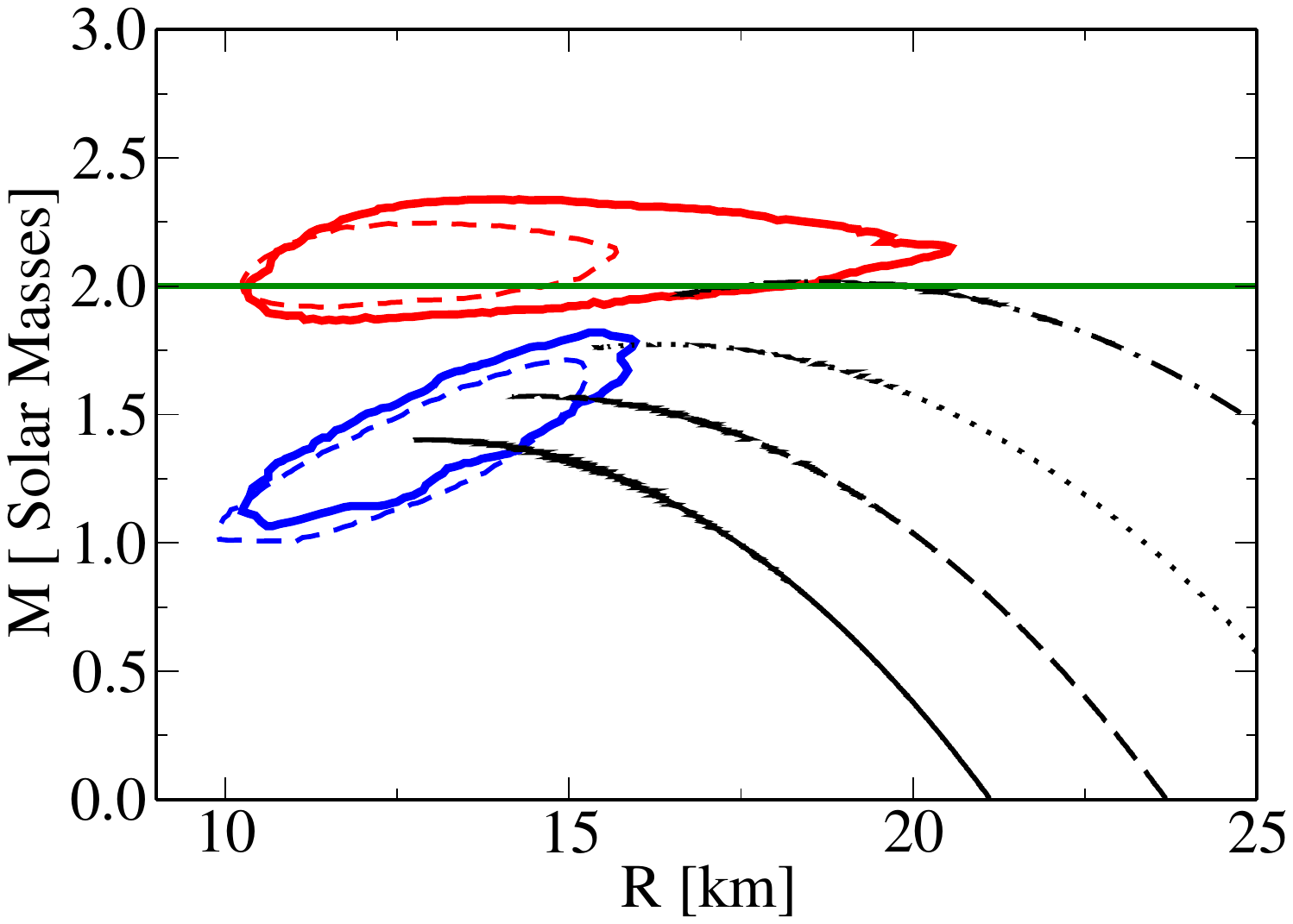}
\end{center}
\vspace{-0.5cm}
\caption{(Color online) QS mass as a function of its radius for different values of $m$. Dot-dashed line: $m = 300$ MeV. Dotted line: $m = 320$ MeV. Dashed line: $m = 340$ MeV. Solid line: $m = 360$ MeV. Red and blue regions represent 95\% confidence intervals for the masses and radii PSR J0030+0451 and PSR J0740+6620 measured by NICER \cite{Riley:2019yda,Miller:2019cac,Miller:2021qha,Riley:2021pdl}. The green horizontal line includes all observed masses over $2 M_\odot$, including the pulsars PSR J1614-2230, PSR J0348+0432
and PSR J0740+6620 \cite{Demorest:2010bx,Antoniadis:2013pzd,NANOGrav:2019jur}.} 
\label{massradius}
\end{figure}

Fig. \ref{massradius} shows the mass-radius sequences of QS using the $D_3-D_7$ holographic EOS. Each sequence of stars was obtained with a particular value of the constituent quark mass, ranging from $m=300$ MeV to $m=360$ MeV. In this figure is clear that increasing the constituent quark mass makes the value of the maximum stellar mass decrease. 

In addition, our computations have been compared with recent observational data analysis from NICER. The millisecond pulsars considered are the PSR J$0030+0451$ \cite{Riley:2019yda,Miller:2019cac} and PSR J$0740+6620$ \cite{Miller:2021qha,Riley:2021pdl}. Independent analysis for PSR J$0030+0451$ gives the inferred mass of $1.34^{+0.15}_{-0.16} M_{\odot}$\cite{Riley:2019yda} and $1.44^{+0.15}_{-0.14} M_{\odot}$ \cite{Miller:2019cac}, while the  radius estimates are 
$12.71^{+1.14}_{-1.19}$ km \cite{Riley:2019yda} and $13.02^{+1.24}_{-1.06}$ km \cite{Miller:2019cac}. For PSR J$0030+0451$, NICER reported the value of $2.072^{+0.0067}_{-0.066} M_{\odot}$\cite{Riley:2021pdl} for the mass, while the radius estimates are 
$13.7^{+2.6}_{-1.5}$ km \cite{Miller:2021qha} and $12.39^{+1.30}_{-0.98}$ km \cite{Riley:2021pdl}. Those range of values are represented by the blue (PSR J$0030+0451$) and red (PSR J$0030+0451$) regions of Fig. \ref{massradius}.

\begin{figure}
\begin{center}
\includegraphics[width=8.5cm, angle=0]{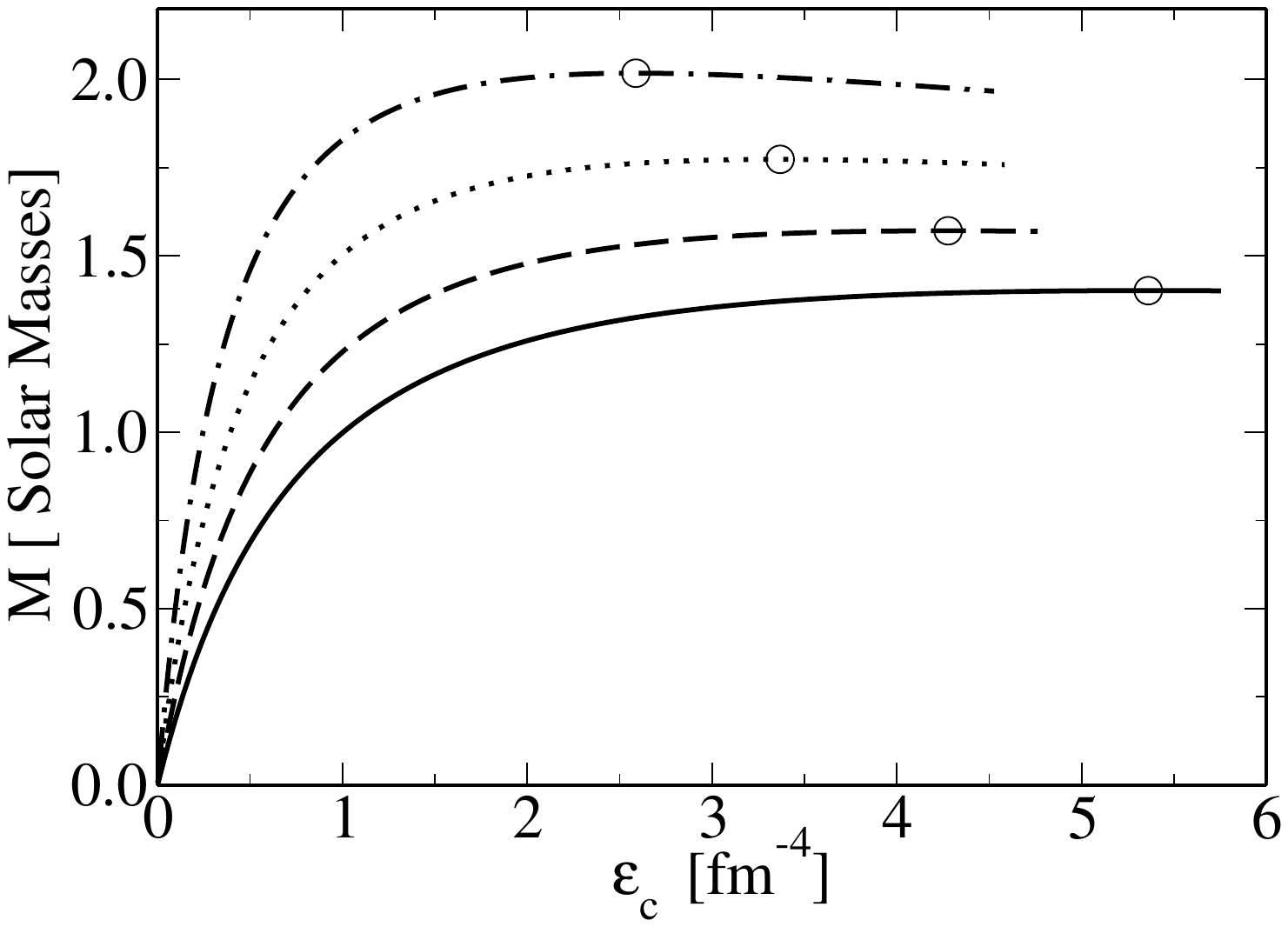}
\end{center}
%\vspace{-0.5cm}
\caption{QS mass $M$ versus central density $\epsilon_c$ for different values of $m$. The maximum mass for each parametrization is shown by a circle. Dot-dashed line: $m = 300$ MeV. Dotted line: $m = 320$ MeV. Dashed line: $m = 340$ MeV. Solid line: $m = 360$ MeV.} 
\label{massdensity}

\end{figure}

The region of stability of the compact stars sequence can be obtained from Fig. \ref{massdensity}. The maximum mass for each parametrization is shown by a circle. All the stars to the left of this point are stable, since $\frac{\partial M}{\partial \varepsilon_{c}}   >  \, 0$ \cite{Shapiro}. Here the static stability criterion is employed, as long as the compact stars under consideration have only one phase.

\begin{figure}[h]
\begin{center}
\includegraphics[width=8.5cm, angle=0]{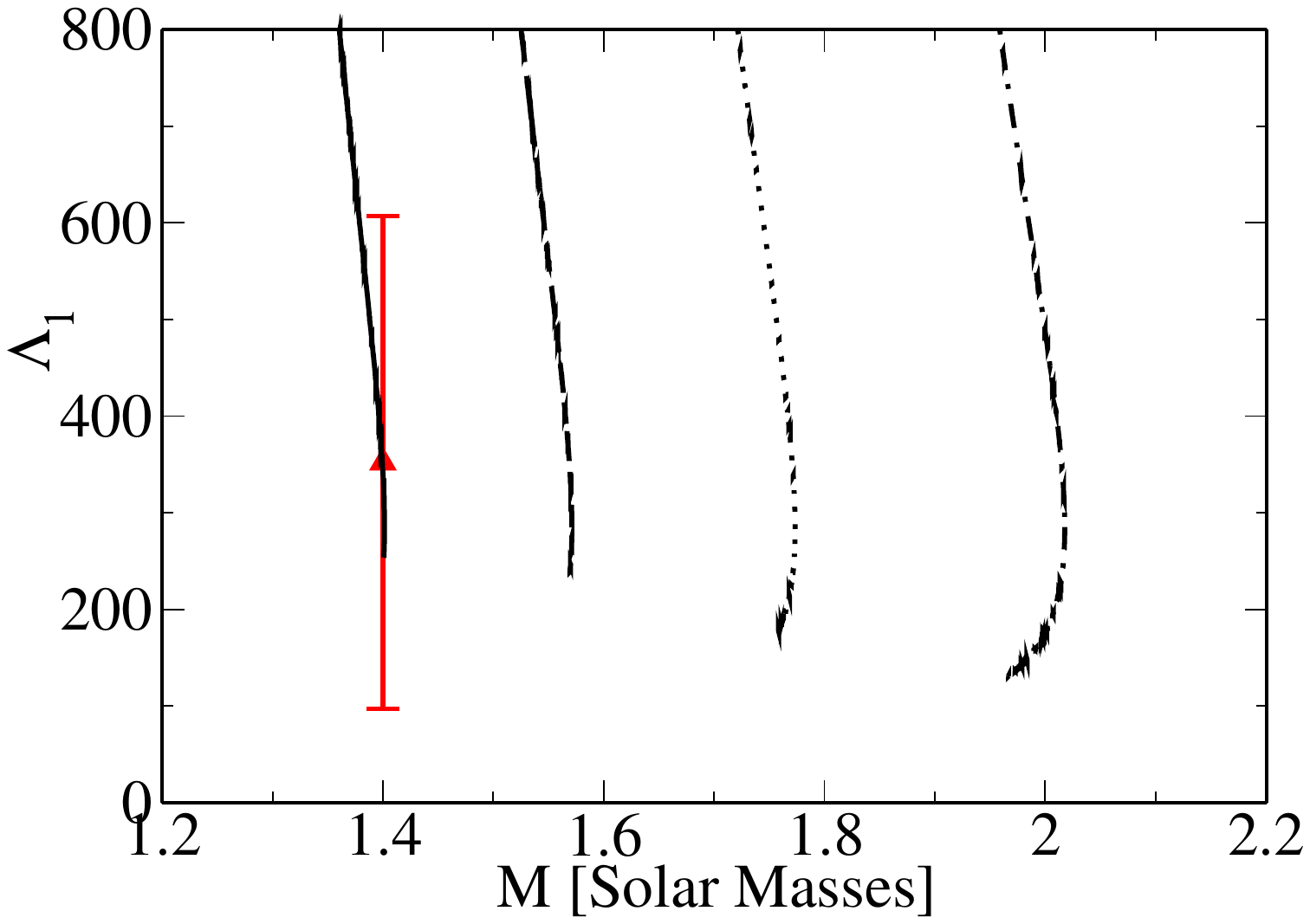}
\end{center}
%\vspace{-0.5cm}
\caption{(Color online) The tidal deformability parameter for the heaviest companion of the NS binary system versus the total stellar mass for different values of $m$. Dot-dashed line: $m = 300$ MeV. Dotted line: $m = 320$ MeV. Dashed line: $m = 340$ MeV. Solid line: $m = 360$ MeV. Observational data from GW170817 event \cite{LIGOScientific:2017vwq,LIGOScientific:2018cki,LIGOScientific:2018mvr}.} 
\label{tidal1}
\end{figure}

For each parametrization, one can solve the TOV equations taking into account the holographic EOS. We use those solutions for $\epsilon(r)$ and $p(r)$ to calculate the relativistic tidal deformability. For this end, we use Eqs. (\ref{lambda}) and (\ref{k2}), performing the integration from the center ($r = 0$) to the star's surface ($r = R$). The outcomes are represented in Fig. \ref{tidal1}. For the constituent quark masses of $360$ MeV, although the tidal deformability obtained is consistent with the GW170817 event, the corresponding maximum mass does not achieve the values expected from NICER observations.

\begin{figure}[t]
\begin{center}
\includegraphics[width=8.5cm, angle=0]{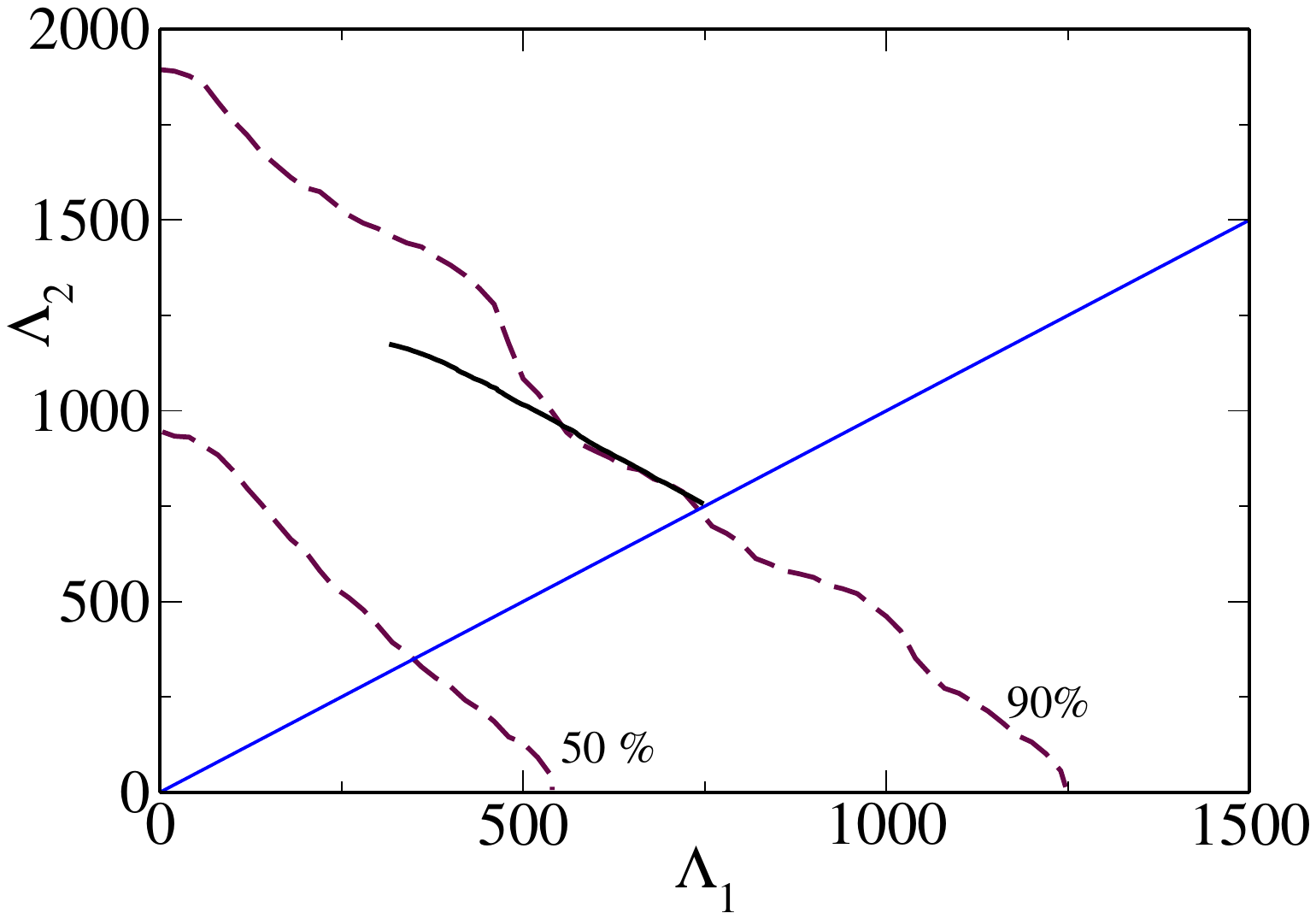}
\end{center}
%\vspace{-0.5cm}
\caption{(Color online) Dimensionless tidal deformability for each component of the binary compact star mergers, GW170817 event, for a constituent quark mass of $360$ MeV within a holographic quark matter description (black solid line). The dashed brown curves correspond to the LIGO-Virgo confidence curves of 
$50\%$ and $90\%$ levels in the low-spin prior scenario \cite{LIGOScientific:2017vwq}, and the blue line indicates the $\Lambda_1 = \Lambda_2$ boundary.
} 
\label{tidal2}
\end{figure}

Fig. \ref{tidal2} presents the dimensionless tidal deformability parameters, $\Lambda_1$ - $\Lambda_2$, for the components of the binary compact star mergers, obtained with the chirp mass of the GW170817 event, $\mathcal{M} = 1.188^{+0.004}_{-0.002} M_{\odot}$. The outcomes for m = $360$ MeV are compared against the LIGO-Virgo confidence curves of 
$50\%$ and $90\%$ levels in the low-spin prior scenario \cite{LIGOScientific:2017vwq}.

It is worth mentioning that the analysis of the model in terms of a range of constituent quark mass $m$ relies on using perturbative QCD results for fixing the value of the 't Hooft coupling constant $\lambda$. Alternatively, if one does not use this large density constrain, the EoS can be written as:
\begin{equation}
\varepsilon = 3 p + \frac{6 \, m^{2} }{\sqrt{\lambda \, \gamma^3}} \sqrt{p} \, \, ,
\label{energydensity2}
\end{equation}
where $\gamma = \Gamma(7/6) \, \Gamma(1/3)/ \sqrt{\pi}$. Note that the parameter $m$ can be rescaled by varying $\lambda$. Therefore, the effect of changing $m$ is the same as a variation of $\lambda$.


\section{Summary and Concluding Remarks} 
In this work, we analyzed both static and dynamical QS properties within a holographic description. The mass-radius relation and the tidal deformability parameter were compared against recent observational data. We solved the TOV equations using the EOS of the $D_3-D_7$ holographic model for describing the quark matter. In this framework, one has $AdS_5 \times S^5$ with the $N_f$ $D_7$-branes wrapping $AdS_5 \times S^3$ \cite{Karch:2007br} and the constituent quark mass is the only adjusted parameter of the EOS. We study the properties of the system for a range of values from $m=300$ MeV to $m=360$ MeV.

We obtained the $M(R)$ sequence of compact stars, highlighting the regions of stability, see Fig. \ref{massdensity}. Decreasing the constituent quark mass value gives a higher maximum stellar mass, the last stable compact star. On the other hand, as the maximum stellar mass increases, the values of the deformability parameter become increasingly far away from the one associated with the GW170817 event observed by the LIGO-Virgo collaboration (see Fig. \ref{tidal1}).

In conclusion, our exploratory study suggests that QS within this holographic model is not able to reproduce simultaneously the tidal deformability of GW170817 event and a stellar mass of $2 M_\odot$. It indicates that further improvements should be implemented as, for example, considering the breaking of the SU(3) flavor symmetry leading to different masses for the quarks up, down and strange. In this case, electrons shall be present to fulfill the beta equilibrium and electric charge neutrality.

QS can describe realistic astrophysical objects, whose quarkyonic matter in the core may carry effects of quantum gravity in AdS/CFT, as reported in Ref. \cite{Kuntz:2022kcw}. The conformal traceless tensor fields, the decay rate of sound waves, the bulk viscosity, the pressure, and the energy density of the QGP were shown to support meaningful quantum corrections due to a functional measure, also encoding the instability of the QGP. Within this framework, the results in Secs. \ref{hcs} -- \ref{res1}  may be  slightly refined when very high-energy processes set in, making the thermodynamic variables acquire these quantum gravity effects.
For instance,  quantum gravity effects account for Eq. (\ref{TOV1}) in Sec. \ref{hcs} and the functions $F(r)$ and $G(r)$ in Sec. \ref{hcs1} to be corrected up to $\sim0.86\%$, when compared to the standard QS without quantum gravity corrections in AdS/CFT. These effects will not significantly change the results obtained in our work, on the scale of energy here studied. Finally,  the stability of  QS, in particular displayed in Fig. \ref{massdensity}, can be alternatively probed by information entropy methods, including the configurational entropy \cite{daRocha:2021jzn,Casadio:2022pla} and the holographic entanglement entropy in QCD \cite{daRocha:2021xwq}.

{\it The authors thank Niko Jokela and Carlos Hoyos for fruitful discussions. M.A. acknowledges the partial support of the National Council for Scientific and Technological Development CNPq (Grant No. 400879/2019-0). C. H. Lenzi is thankful to the S\~ao Paulo Research Foundation FAPESP (Grant No. 2020/05238-9) and to CNPq (Grants No. 401565/2023-8 and 305327/2023-2). W.d.P. acknowledges the partial support of CNPq (Grant No. 313030/2021-9) and the Coordination for the Improvement of Higher Education Personnel CAPES (Grant No. 88881.309870/2018-01). R.d.R.~is grateful to FAPESP (Grant No. 2021/01089-1 and No. 2022/01734-7), CNPq (Grant No. 303390/2019-0), and CAPES-PrInt (Grant No. 88887.897177/2023-00), for partial financial support; and to Prof. Jorge Noronha and the Illinois Center for Advanced Studies of the Universe, University of Illinois at Urbana-Champaign, for the hospitality.
}

\end{document}